\documentclass[letterpaper, 10 pt, conference]{ieeeconf}  

\IEEEoverridecommandlockouts                              

\usepackage{graphics} 
\usepackage{graphicx}
\usepackage{multirow}
\usepackage[spaces,hyphens]{url}

\title{\LARGE \bf
Automatic Classification of OSA related Snoring Signals from Nocturnal Audio Recordings
}

\author{Arun Sebastian \textit{Student Member, IEEE}, Peter A. Cistulli,  Gary Cohen and Philip de Chazal \textit{Senior Member, IEEE}  
\thanks{Research supported by University of Sydney}
\thanks{Arun Sebastian and Philip de Chazal are with the Sleep Research Group, Charles Perkins Centre, School of Biomedical Engineering, University of Sydney, Sydney, NSW, 2006, Australia
        {\tt\small \{arun.sebastian,philip.dechazal\}@sydney.edu.au}}%
\thanks{Peter A. Cistulli is with the Charles Perkins Centre and Northern Clinical School, University of Sydney, Sydney, NSW, 2006, Australia
        {\tt\small peter.cistulli@sydney.edu.au}}%
\thanks{Gary Cohen is with the Sleep Investigation Laboratory, Center for Sleep Health and Research, Royal North Shore Hospital, Sydney, NSW, 2065, Australia
        {\tt\small Gary.Cohen@health.nsw.gov.au}}%
}

\begin{document}

\maketitle
\thispagestyle{empty}
\pagestyle{empty}

\begin{abstract}

In this study, the development of an automatic algorithm is presented to classify the nocturnal  audio recording of an obstructive sleep apnoea (OSA) patient as OSA related snore, simple snore and other sounds. Recent studies  has been shown that knowledge regarding the OSA related snore could assist in identifying the  site of airway collapse. Audio signal was recorded simultaneously with full-night polysomnography during sleep with a ceiling microphone. Time and frequency features of the nocturnal audio signal were extracted to classify the audio signal into OSA related snore, simple snore and other sounds. Two algorithms were developed to extract OSA related snore using an linear discriminant analysis (LDA) classifier based on the hypothesis that OSA related snoring can assist in identifying the site-of-upper airway collapse. An unbiased  nested leave-one patient-out cross-validation process was used to select a high performing feature set from the full set of features. Results indicated that the algorithm achieved an accuracy of 87\% for identifying snore events from the audio recordings and an accuracy of 72\% for identifying OSA related snore events from the snore events. The direct method to extract OSA-related snore events using a multi-class LDA classifier achieved an accuracy of 64\% using the feature selection algorithm. 
Our results gives a clear indication that OSA-related snore events can be extracted from nocturnal sound recordings, 
and therefore could potentially be used as a new tool for identifying the  site of airway collapse from the nocturnal audio recordings.

\end{abstract}

\section{INTRODUCTION}

The gold-standard method for the diagnosis of OSA is  Polysomnography (PSG), where the patient’s sleep is monitored overnight by recording a suite of  physiological signals [1]. The disadvantages of the overnight PSG in an attended centre include the expense of the test, long waiting times for patients due to limited resources, and poor sleep quality of patients  during the study due to the attachment of uncomfortable sensors.    As snoring data can be  conveniently recorded using simple and non-invasive techniques, acoustic features extracted from snoring have been successfully implemented in the diagnosis of OSA and to identify OSA severity.

Several studies have been conducted in developing an automatic snore signal extraction algorithms from nocturnal audio recording  with a high accuracy. Microphone data was used  to classify  audio signals as a snoring episode or a non-snoring episode using  machine learning models such as Hidden Markov  [2],  fuzzy C-means clustering   [3], $k$-means clustering  [4],  artificial neural network (ANN)  [5], and recurrent neural network  [6]. Studies have also been conducted to identify  correlation between  acoustic features of snoring and OSA [7-9]. One study proposes a novel feature of snore signal termed the 'intra-snore-pitch-jump'  to diagnose OSA with sensitivities of 86-100\% while holding specificity at 50-80\% [10]. In another study  snore signal was adopted as a screening tool for the diagnosis of OSA by extracting low-level acoustic features  using semi-automatic algorithm based on Gaussian mixture models  and achieved correct classification of 92\% for resubstitution method and 80\% for 5-fold cross validation method [11].  An OSA severity identification algorithm was developed using an inter-(apnoea phase ratio, running variance,  inter-event silence)snore property and a Bayes classifier, and the audio was recorded using a non-contact microphone [12]. This method achieved area under the receiver operating characteristic curve of 85\% and 92\% for thresholds of 10 and 20 events/h, respectively, were obtained for OSA detection. Alencar et al. observed  the correlation between number of irregular snores  with AHI, which was identified using Hurst analysis of the snore sound [13].   A two-layer neural network was developed to automatically detect snore from a  full-night audio recording using a tracheal microphone and to classify snorers based on OSA severity [14]. 

Our previous studies have demonstrated that snore during OSA  events can be classified based on the  site-of-
collapse as lateral wall, palate and tongue-based related collapse using an automatic classifier [15, 16]. This study was extended  to label OSA patients into 4 categories (``lateral wall", ``palate", ``tongue-base" related collapse and ``multi-level" site-of-collapse) based on the predominant site-of-collapse of the hypopnoea events of a night's recording using snore data, and achieved an accuracy of 81\%[17]. An LDA classifier [15-18] and $k$-means clustering model [19] were developed to identify the predominant site-of-collapse.  Extracting OSA related snore signal was the first important step in these studies.  For the above mentioned studies, OSA events were first manually identified before OSA related snore signal could be extracted, which is an major drawback. Therefore, the  objective of this study is to develop a classification model that can automatically extract OSA related snore from the nocturnal audio recordings.

Researchers have previously used a two-step process to identify apnoea events for OSA diagnosis and to identify its severity [7-14], as shown in Figure 1. The first step is a snore detector, which extracts the snore signal from the audio recording. Snore events are then further processed to determine if they constitute an apnoea or hypopnoea event-related snore or simple snore. Using the apnoea or hypopnoea event information, OSA diagnosis and OSA severity classifications are made. In all of these studies, the performance evaluation was based on comparing the model with AHI, which was evaluated using PSG data.

   \begin{figure}[h!]
      \centering

\begin{center}
\underline{}
\end{center}     \includegraphics[scale=.31]{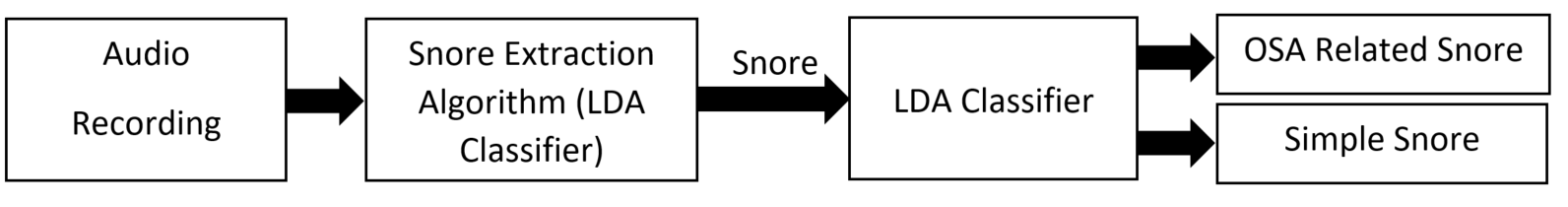}
      \caption{Block diagram of the conventional two-step process to extract OSA related snore. 
       }
      \label{figurelabel}
   \end{figure}

   \begin{figure}[h!]
      \centering

\begin{center}
\underline{}
\end{center}     \includegraphics[scale=.38]{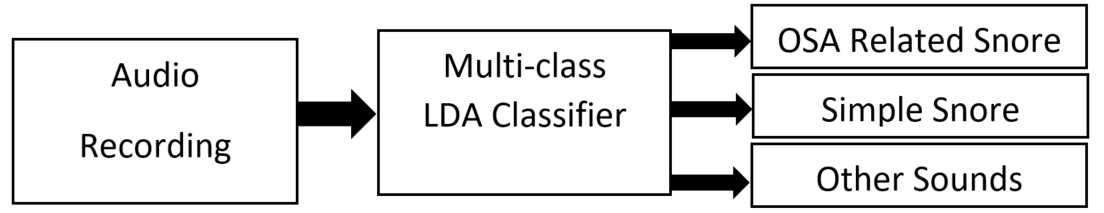}
      \caption{Block diagram of the direct method to extract OSA related snore.  
       }
      \label{figurelabel}
   \end{figure}
   
For the current study, the aim was to extract snoring during hypopnoea events, which could be beneficial in the identification of the site-of-collapse without using invasive technology.  For the extraction of the OSA-related snore, two methods were adopted. The first method adopted the conventional two-step process by identifying OSA-related snoring from the snore detection algorithm, as shown in Figure 1. In the second method, an automatic system using machine learning algorithm was implemented to directly classify the audio signal recorded using a full-night sleep study into OSA-related snore (snoring during a hypopnoea event), simple snore (snoring episode not associated with an apnoea event) and other sounds (normal breathing, noise from body movement, coughing, talking and other environmental noises), as shown in Figure 2. This classification was carried out using an LDA classifier and  the most relevant features for the model were identified using a nested cross-validation technique.



\section{METHOD}

\subsection{Data Collection}

The data from  58 patients  who  attended for a full night sleep study using PSG at the Sleep Investigation Unit, Royal North Shore Hospital, Sydney, were used for this study. The audio signal were recorded along with PSG during sleep  with a microphone, which is placed on the ceiling about 1.5$m$ above the patient's bed.  Audio data was sampled  16kHz and were synchronized with the PSG signal. This study received ethical approval by the Northern Sydney Local Health District Human Research Ethical Committee as application RESP/18/184.

\subsection{Data Labelling }
Sleep stages and respiratory events annotations were collected along with the PSG signal and the audio recordings.  For the classification of the audio signal into OSA related snore, simple snore (other than snoring during  hypopnoeas) and non-snore, we manually labelled the audio recording. Manual labelling of audio recording was based on visual and auditory inspection of the audio recordings. Assistance was provided by a sleep expert to identify the snore events as there was no definitive definition for a ``snoring event”. For this study, OSA related snoring was defined as the snoring during an hypopnoea event, and this was manually labelled and extracted with the help of PSG annotations. Simple snoring was  defined as  snoring that did not occur during a hypopnoea event. Simple snoring was labelled and extracted if a clear acoustic perception of snoring was apparent from the audio recording. Non-snore episodes were defined as episodes  of other sounds such as normal breathing, noise from body movement, coughing, talking and other environmental noises. A typical example of labelling simple-snore and OSA related snore for an audio recording is shown in Figure 3. By this process,  we have a database  from  58 patients with 2666 hypopnoea events with snoring (average of 20 minutes per patient, ranging from 7 minutes to 50 minutes), simple snore (average of 1hr 45 minutes for each patient, ranging from 30 minutes to 3hrs. 30 minutes) and other sounds (average of 3hrs. 30 minutes for each patient ranging from 2hrs. 30 minutes to 4 hours 45 minutes).

    \begin{figure}[h!]
       \centering
 
 \begin{center}
 \underline{}
 \end{center}     \includegraphics[scale=.2]{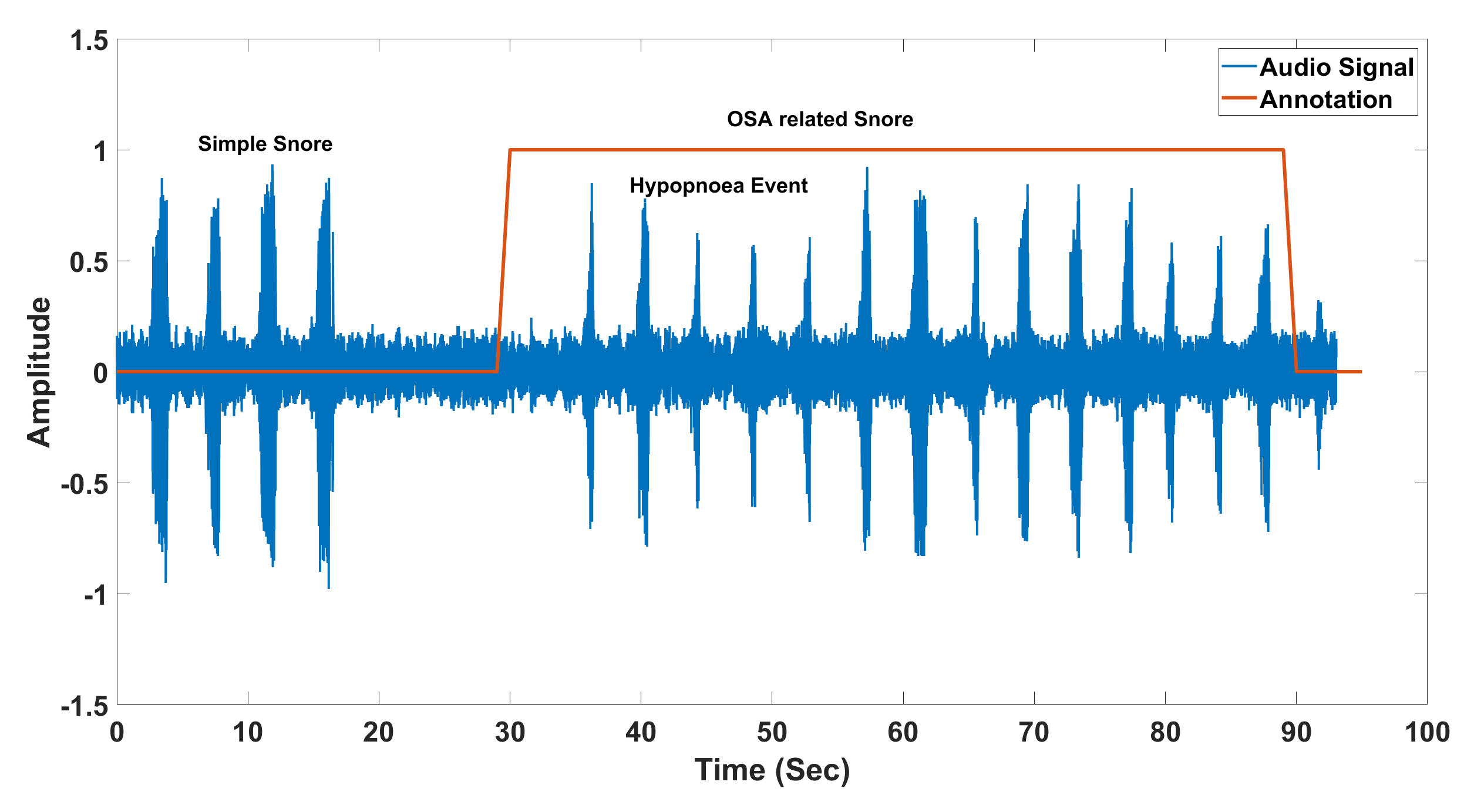}
       \caption{An example of labelling audio recording (90s epoch) corresponding to simple-sore and OSA related snore.}
       \label{figurelabel}
    \end{figure}

\subsection{Signal Processing}
\subsubsection{Preprocessing}
The audio signal was recorded using a general-purpose microphone   placed 1.5m above the patient. As the raw audio signal contained significant background noise (noise from an air conditioner, hiss and hum), enhancement of raw signal was performed to improve the signal to noise ratio (SNR). We trailed three methods of noise reduction techniques, including spectral subtraction [19], multi-band spectral subtraction algorithm [21] and a band-pass filter method and we found that spectral subtraction was the most effective method for the current application.
 
\subsubsection{Feature Extraction}

We extracted 50 identical features derived from the audio recording, consisting of time and spectral
features. Details are described in [17] and we briefly summarise them here. Time-domain features consisted of (1) Energy (2) Entropy (3) ZCR. The frequency-domain feature consisted of (1) First three formant frequencies (2) Thirteen MFCC and its first derivative (3) Twelve spectral chroma features (4) Spectral Entropy, Flux, Centroids and roll-off (5) Fundamental Frequency and Harmonic Frequency.  Feature extraction was done using the window of width 10s without overlapping. Based on this feature extraction framework, features from approximately 50,000 audio events ($\sim$30,000 other sound events; $\sim$17,000, simple snore event; $\sim$3000 OSA related snore events) were extracted and used for this study. We used an unbiased performance estimation process  using nested leave-one patient-out cross-validation to choose a high performing subset of the available features that best predicted the site-of-collapse.

\subsection{Machine Learning}
\subsubsection{LDA Classifier}
The Linear Discriminant Analysis (LDA) algorithm is one of the simplest and most widely used  classification algorithms. The LDA classifier projects the data from a high dimensional space into a lower dimension by maximising the separability between different classes of events with linear boundaries. It is  computed by calculating the variance of data within and between classes [22].  The LDA models each class by a Gaussian distribution. It assumes the  features are statistically independent, and that the data has the same covariance for all classes. Classification is achieved through a simple probabilistic decision-making algorithm using the maximum likelihood principle which models the conditional distribution of the data.
\subsubsection{Nested Cross Validation}
Nested cross-validation is the most popular way to independently select the best parameters to train an optimal prediction model and get an  unbiased  estimate of its performance [23]. Nested cross validation comprises of double cross-validation loops, and it is performed to obtain a performance estimation on the training set to find the optimal hyper-parameters for the model. Training data from the outer loop is split into 10-folds to create an inner fold, as shown in Figure 4. The outer loop of cross-validation is used to provide the performance estimate, and the inner-loop used to select and tune the hyper-parameters of the model. Once the optimal hyper-parameters are chosen, the classification performance is evaluated on the test data. 
   \begin{figure}[h!]
      \centering

\begin{center}
\underline{}
\end{center}     \includegraphics[scale=.28]{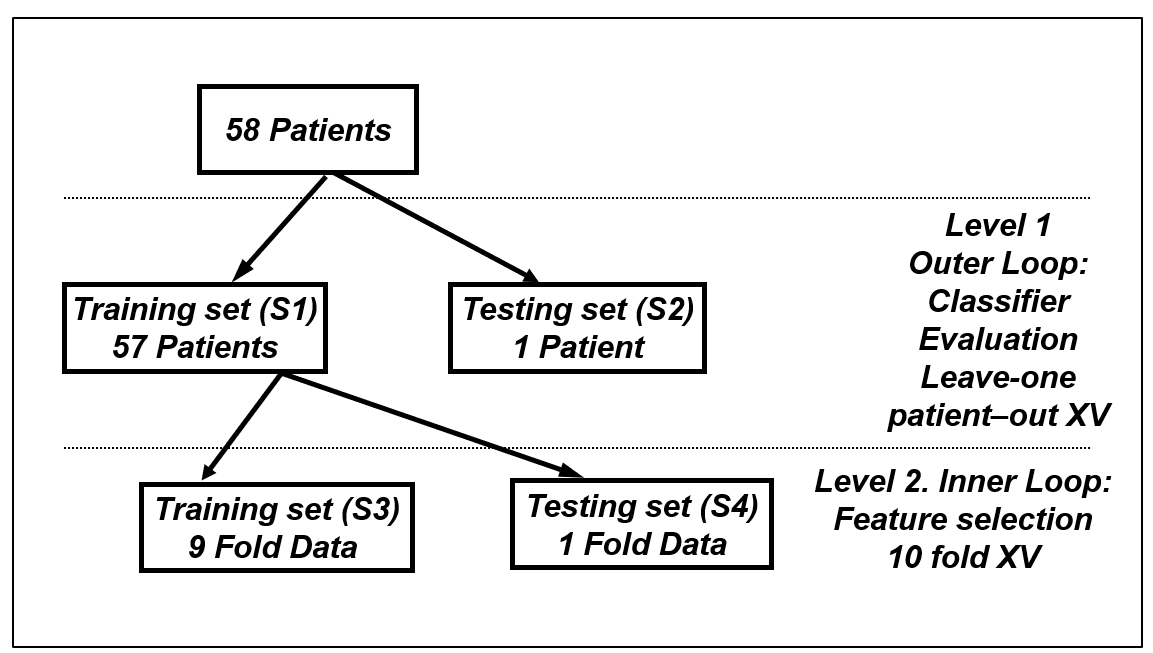}
      \caption[Data partitioning for the model]{Data partitioning for the model. Three levels of data partitioning resulting
            in four datasets (S1-S4) were used to provide unbiased results. 
       }
      \label{figurelabel}
   \end{figure}

\section{RESULTS }

The LDA classifier was deployed to extract OSA related snore events from a full night nocturnal  audio recording. A nested-leave-one-patient-out cross-validation method was adopted to identify the best model with the most relevant features for each test subjects. Two algorithms were developed to extract OSA related snore from the audio recording. To evaluate the performance of OSA related snoring extraction, we  performed three experiments using the LDA classifier, as follows:

\subsection{Identification of Snore Events From Audio Recordings}
In the first phase, the LDA classifier was deployed to classify the full night audio recording as snore events and other sounds.  Snore events consisted of all  OSA related  snore events and all other  snoring episodes. The classification performance was evaluated using all of the features extracted and the most relevant features identified using the nested cross-validation method.   Classification results showed that the LDA classifier achieved an overall accuracy of 82\% (95\% CI, 80\%-84\%) for classifying snore events and other sounds using all of the available features. The detailed results are shown in Table I. The classification results achieved an overall accuracy of 87\% (95\% CI, 86.6\% - 87.3\%) when classifying snore events and other sounds using the selected features from the nested cross-validation technique. Results from the feature selection outperformed the results using all of the features (see Table I). Evaluated across the different cross-validation data splits, there was an average of 19 features selected, which is less than 40\% of total the features extracted. The most frequently selected  features were the  first derivative MFCC coefficients,  MFCC coefficients, fundamental and formant frequency,  spectral chroma features, spectral entropy, and energy. The main reason for the low positive predictive values (PPV) may be due to the low number of snore events compared with the other sound events, which can  introduce a small bias towards the other sound events. 

 \renewcommand{\arraystretch}{1.5}
 \begin{table}[h]
 \caption{Cross-validation results for snore and non-snore classification} \vspace{3mm}
 \centering
\begin{tabular}{c|c|c|}
\cline{2-3}
                                                         & \textbf{All features}   & \textbf{Using feature selection} \\ \hline
\multicolumn{1}{|c|}{\textbf{Statistic}}                 & \textbf{Value (\% with 95\% CI)} & \textbf{Value (\% with 95\% CI) }         \\ \hline
\multicolumn{1}{|c|}{\textbf{Accuracy}}               & 82 (80 - 84)            & 87 (86.6 – 87.3)                 \\ \hline
\multicolumn{1}{|c|}{\textbf{Sensitivity}}               & 86 (85 - 87)            & 90 (89.5 – 90.5)                 \\ \hline
\multicolumn{1}{|c|}{\textbf{Specificity}}               & 81 (80 - 81.5)          & 85.5 (85 - 86)                   \\ \hline
\multicolumn{1}{|c|}{\textbf{PPV}} & 66 (65 – 67)            & 75 (74 – 75.5)                   \\ \hline
\multicolumn{1}{|c|}{\textbf{NPV}} & 93 (92.5 - 93.5)        & 95 (94.5 - 95.5)                 \\ \hline
\end{tabular}
\end{table}

\subsection{Extraction of OSA Related Snoring From All Snore Events}
Once the snore event had been identified, the next phase was to extract the OSA related snoring from the snore events. For this  phase, the LDA classifier was deployed to classify snore events as OSA related snore and simple snore using all the features extracted as well as the most relevant features using the nested cross-validation method.  For this experiment, an equal number of OSA related snore and simple snore events were selected for each patient to obtain balanced data for the classification model. 

The classification results showed that the LDA classifier achieved an overall accuracy of 67\% (95\% CI, 66\%-69\%) when classifying snore events into OSA snore events and simple snore using all of the features and with an  accuracy of 72\% (95\% CI, 71\%-74\%) using the selected features from the nested cross-validation technique.  The detailed results are shown in Table II. Evaluated across the different cross-validation data splits, there was an average of eight features selected, which is less than 20\% of the total features extracted.  The most commonly selected  features were energy, ZCR,  MFCC coefficients,  first derivative of MFCC coefficient,  formant frequency,  spectral chroma feature and spectral entropy. Comparing the performance of the feature selection  using all of the features illustrated that feature selection  resulted in a higher-performing model.

 \renewcommand{\arraystretch}{1.5}
 \begin{table}[]
 \caption{Cross-validation results for the extraction of OSA related snoring} \vspace{3mm}
 \centering
\begin{tabular}{c|c|c|}
\cline{2-3}
\textbf{}                                                & \textbf{All features}            & \textbf{Using feature selection} \\ \hline
\multicolumn{1}{|c|}{\textbf{Statistic}}                 & \textbf{Value (\% with 95\% CI)} & \textbf{Value (\% with 95\% CI)} \\ \hline
\multicolumn{1}{|c|}{\textbf{Accuracy}}               & 67 (66 - 69)            & 72 (71 – 74)                 \\ \hline
\multicolumn{1}{|c|}{\textbf{Sensitivity}}               & 69 (67 - 71)                     & 73 (71 – 74.5)                   \\ \hline
\multicolumn{1}{|c|}{\textbf{Specificity}}               & 65 (63.5 - 67.5)                 & 72 (70 - 74)                     \\ \hline
\multicolumn{1}{|c|}{\textbf{PPV}} & 69.5 (68 – 71)                   & 77 (75.5 – 78)                   \\ \hline
\multicolumn{1}{|c|}{\textbf{NPV}} & 65 (63 - 67)                     & 67 (66 - 69)                     \\ \hline
\end{tabular}
\end{table}

\subsection{Direct Method to Extract OSA Related Snore Events From  Audio Recordings}
In this experiment, a multi-class LDA classifier was developed to classify the nocturnal audio recording into OSA related snore, simple snore and other sounds. This method is capable of directly extracting OSA related snore from audio recordings without using the conventional two-step process. The classification was performed using all of the features extracted, and the most relevant features were extracted using the nested cross-validation method. The classification results showed that the LDA classifier achieved an overall accuracy of 59\% (95\% CI, 57\%-61\%) when classifying the snore events into OSA snore events and simple snore using all of the features, and with an accuracy of 63\% (95\% CI, 61\%-65\%) using the selected features from the nested cross-validation technique.   The detailed results are shown in Table III. The results demonstrated that the feature selection method using nested cross-validation performed better than using all of the features. Evaluated across the different cross-validation data splits, there was an average of 18 features selected, which is less than 40\% of the total features extracted.  The most commonly selected 18 features were energy, entropy, ZCR, spectral entropy, four MFCC coefficients, five spectral chroma features, 2nd and 3rd formant frequencies and three first derivative MFCC coefficients.
 \begin{table}[]
 \caption{Cross-validation results for the direct method to extract OSA related snore events   } \vspace{3mm}
 \centering
\begin{tabular}{c|c|c|c|}
\cline{2-4}
                                                                                                            & \textbf{Parameters}  & \textbf{\begin{tabular}[c]{@{}c@{}}All features\\ (\% with 95\% CI)\end{tabular}} & \textbf{\begin{tabular}[c]{@{}c@{}} Feature selection\\ (\% with 95\% CI)\end{tabular}} \\ \cline{2-4} 
                                                                                                            & \textbf{Accuracy}    & 59 (57 - 61)                                                                      & 63 (61 - 63)                                                                                 \\ \hline
\multicolumn{1}{|c|}{\multirow{2}{*}{\textbf{\begin{tabular}[c]{@{}c@{}}OSA Related\\ Snore\end{tabular}}}} & \textbf{Sensitivity} & 60 (58 - 62)                                                                      & 65 (63 - 67)                                                                                 \\ \cline{2-4} 
\multicolumn{1}{|c|}{}                                                                                      & \textbf{PPV}         & 66 (64 - 67.5)                                                                    & 69 (68 - 70)                                                                                 \\ \hline
\multicolumn{1}{|c|}{\multirow{2}{*}{\textbf{Simple Snore}}}                                                & \textbf{Sensitivity} & 60 (59 - 61)                                                                      & 64 (62 - 67)                                                                                 \\ \cline{2-4} 
\multicolumn{1}{|c|}{}                                                                                      & \textbf{PPV}         & 49 (47 - 50)                                                                      & 52 (49 - 53)                                                                                 \\ \hline
\multicolumn{1}{|c|}{\multirow{2}{*}{\textbf{Other Sound}}}                                                 & \textbf{Sensitivity} & 61 (59 - 63)                                                                      & 63 (61 - 65)                                                                                 \\ \cline{2-4} 
\multicolumn{1}{|c|}{}                                                                                      & \textbf{PPV}         & 67 (65 - 67)                                                                      & 71 (69 - 73)                                                                                 \\ \hline
\end{tabular}
\end{table}
\section{DISCUSSION}

 \renewcommand{\arraystretch}{1.5}
 \begin{table*}[h!]
 \caption{Classification performance comparison for snore and non-snore events } \vspace{3mm}
 \centering
\begin{tabular}{|c|c|c|c|}
\hline
 \textbf{Method}              & \textbf{Subjects}        & \textbf{\begin{tabular}[c]{@{}c@{}}Microphone placement\\ (Distance from head:\textit{cm})\end{tabular}} & \textbf{Accuracy (\%)} \\ \hline
Hidden Markov Model [2]          & OSA patients (6)         & 20                                                                                              & 82-89                  \\ \hline
 Linear Regression [24]            & SS and OSA patients (30) & 15                                                                                              & 87                     \\ \hline
 fuzzy C-means clustering [3]    & OSA patients (30)        & 20-30                                                                                           & 93                     \\ \hline
 ANN classifier [5]               & OSA patients (34)        & 40-50                                                                                           & 86-89                  \\ \hline
 AIM technique [25]               & SS and OSA patients (40) & 50                                                                                              & 96                     \\ \hline
 Recurrent Neural Network [6]   & SS and OSA patients (40) & 70                                                                                              & 95                     \\ \hline
                \textbf{Our proposed method} & \textbf{OSA patients (58)}    & \textbf{150}                                                                                    & \textbf{89}            \\ \hline
\end{tabular}
\end{table*}
An automatic classification algorithm using an LDA classifier was developed to extract OSA related snore during hypopnoeas from a full night  audio recording. Two methods were adopted to label OSA related snore. The first method adopted a conventional two-step process by identifying OSA related snoring from the snore detection algorithm, while the second method directly extracted OSA related snore using a multi-class LDA classifier. The results indicate that snore events can be extracted from audio recording with an accuracy of 87\% using  selected features and can achieve an accuracy of 72\% when identifying OSA related snore from the snore events. The direct method adopted  to extract OSA related snore events achieved an accuracy of 63\% when using the feature selection algorithm. The results showed that the model achieved considerable accuracy when directly extracting the OSA related snore. Overall, both methods resulted in a similar performance with a slightly higher accuracy for  the direct method  compared with the conventional method. It can also be noted here that the feature selection algorithm resulted in a higher performance across all of the  experiments conducted. The feature selection algorithm indicates that most of the features consisted of frequency-related features across all  experiments (18 out of 19 features in the snore detection algorithm, six out of eight features when identifying OSA related snore, and 15 out of 18 features for the direct method to extract OSA related snore). As MFCC features made up nine of the 18 selected features in the snore detection algorithm, three out of eight features when identifying OSA related snore, and seven out of 18 features for the direct method to extract OSA related snore, our results indicate that MFCC features  provided the most discriminating information. This result is consistent with the observation that MFCCs have  successfully been used by other researchers in OSA diagnosis and  severity classification using audio analysis. 

Table IV shows the classification performance comparison for snore and non-snore events. A study employing the Hidden Markov model and using  spectral-based features to identify snore for a small training set  achieved an accuracy of 82-89\%  [2]. Another study developed a snore-event extraction algorithm based on linear regression that stemmed from  sub-band energy distributions [24]. This study achieved an accuracy of 90\% when the model was trained using  data from  simple snorers and OSA patients. It achieved an accuracy of 87\% for detecting snore episode when trained with OSA patient data. Azarbarzin et al. adopted an unsupervised machine learning technique using a fuzzy C-means clustering algorithm to label the full night audio recording as  snore or no-snore episode, and the audio was recorded using a tracheal microphone [3]. The results showed that the model   achieved an overall accuracy of 98\% and 93\% using tracheal sound recording and microphone recording, respectively to extract snore events from audio recordings. Swarnkar et al.  deployed  an ANN classifier to classify snore and non-snore events from nocturnal sound recordings and achieved an accuracy of 86-89\% [5].   Nonaka et al. achieved a  high accuracy of 96\%  for extracting  snore events   from the audio recordings using the auditory image modelling (AIM) technique [25]. The data set for training consisted of OSA patients and simple snorers. Arsenali et al. achieved an accuracy of 95\% for extracting snore events by an adaptive energy threshold method using data from OSA patients and simple snorers [6]. Our current method showed that snore events could be extracted from OSA patients’ nocturnal sound recordings with an accuracy of  87\%.  The result indicate that our proposed method can achieve  robust results compared to the other published  methods in extracting snore events from audio recordings. Although this study consists of more study participants, the microphone placement was far from patient compared to other studies. The microphone was not intended to record special sounds such as the patient’s breath. We believe that these reasons might result in slightly lower snore detection rate compared to other studies [3, 6, 25].

Although various studies have been conducted to diagnose OSA and its severity classification [11-14], all of these studies were based on estimating the AHI range. To the author’s knowledge, no studies have demonstrated the performance evaluation of the system when identifying OSA-related snore events.  For the current study, the main objective was to extract OSA-related snore during hypopnoea events from nocturnal sound recordings. This was based on the hypothesis that OSA-related snoring can assist in identifying obstructions sites.   Classification based on AHI was beyond the scope of this study. The method proposed in the current study resulted in the achievement of high accuracy when extracting OSA-related snore and gave a clear indication that OSA-related snore events can be extracted from nocturnal sound recordings.

There are several limitations to the current study. First, the audio was recorded using a simple microphone, which was not designed to record breathing sounds, and it was positioned at approximately 1.5$m$ above the patient’s bed. This resulted in poorer sound quality than could have been achieved with a more specialised setup. Furthermore, the audio signal was selected without considering the effect of body position. Therefore, both limitations may have affected the signal quality and audio frequencies of the recording. I expect that a better performance could be achieved with a more specialised setup. Also, further studies are necessary utilising non-linear advanced machine learning algorithms for possible better model performance.

\section{ CONCLUSION }

This study proposes an automatic classification algorithm using an LDA classifier to extract OSA-related snore events from a nocturnal audio recording. The acoustic properties of OSA related snoring could potentially be used as a non-invasive method to identify obstruction sites in OSA patients. An unbiased process using nested leave-one-patient-out cross-validation results showed that the snore events were extracted from audio recordings with an accuracy of 87\% using the selected features and achieved an accuracy of 72\% when identifying OSA-related snore from the snore events. The direct method to extract OSA-related snore events achieved an accuracy of 64\% using the feature selection algorithm. The use of this method achieved a reasonable high accuracy when extracting OSA-related snore and gave a clear indication that OSA-related snore events can be extracted from nocturnal sound recordings.

\addtolength{\textheight}{-12cm}   



\end{document}